# CHROMOSPHERIC PECULIAR OFF-LIMB DYNAMICAL EVENTS FROM *IRIS* OBSERVATIONS

E. Tavabi[1] and S. Koutchmy[2]

[1] *Physics Department, Payame Noor University (PNU), 19395-3697, Tehran, I. R. of Iran.*
[2] *Institut d'Astrophysique de Paris, CNRS and Sorbonne Univ. UPMC, UMR 7095, 98 Bis Bd. Arago, 75014 Paris, France.*

## ABSTRACT

To study motions and oscillations in the solar chromosphere and at the transition region (TR) level we analyze some extreme Doppler shifts observed off-limb with the *Interface Region Imaging Spectrograph (IRIS).* Raster scans and slit-jaw imaging observations performed in the near-ultraviolet (NUV) channels were used. Large transverse oscillations are revealed by the far wings profiles after accurately removing the bulk average line profiles of each sequence. Different regions around the Sun are considered. Accordingly, the cool material of spicules is observed in Mg II lines rather dispersed up to coronal heights. In the quiet Sun and especially in a polar coronal hole, we study dynamical properties of the dispersed spicules-material off-limb using a high spectral, temporal and spatial resolutions *IRIS* observations. We suggest that numerous small-scale jet-like spicules show rapid twisting and swaying motions evidenced by the large distortion and dispersion of the line profiles, including impressive periodic Doppler shifts. Most of these events repeatedly appear in red- and blue-shifts above the limb throughout the whole interval of the observation datasets with an average swaying speed of order $\pm 35$ km s$^{-1}$ reaching a maximum value of 50 km s$^{-1}$ in the polar coronal hole region, well above the 2.2 Mm heights. We identified for the 1$^{st}$ time waves with a short period of order of 100 sec and less and transverse amplitudes of order of $\pm$ 20 to 30 km s$^{-1}$ with the definite signature of Alfven waves. No correlation exists between brightness and Doppler shift variations; the phase speed of the wave is very large and cannot definitely be determined from the spectral features seen along the quasi-radial features. Even shorter periods waves are evidenced, although their contrast is greatly attenuated by the overlapping effects along the line of sight.

Corresponding author: Ehsan Tavabi
etavabi@gmail.com

# 1. INTRODUCTION

The first analysis of the TR and chromospheric spectral lines and structures crossing the solar limb were done several decades ago (e.g. Nikolskii 1965; Zirin 1966; Brueckner and Bartoe 1983) with photographic emulsion as a detector. More recently using the TRACE mission data (Alissandrakis et al. 2005) the EUV time sequences were analyzed. Looking near and above the solar limb have the advantage of straightforwardly providing the geometric heights, which is not warrantied in disk observations, phenomena over a longer integration along the line of sight are analyzed; they preferably show transverse motions when Doppler effects are measured. A great amount of data is made available from IRIS higher resolution spectral and image observations (De Pontieu et al. 2004), so it is timely to carry out again new studies of the transverse and twisty motions of spicule-like features and of TR macrospicules using spectra. IRIS observations provides simultaneous filtergrams and a wide range of spectra made with an unprecedented high resolution 0"33 width slit. This allows us to investigate the velocities in great details. Studies of such motions have been done in several recent papers (e.g. De Pontieu et al. 2014a; Tavabi et al. 2014a, 2015a). Alissandrakis et al. (2018) performed an extended work to deduce the ''average'' properties of a static chromosphere from the IRIS mission FUV data. The motions of spicules were previously less well established from spectra; some evidence of spinning motions were however reported using Solar Optical Telescope onboard the Hinode in Ca II H line imaging observations with swaying velocities of 20 to 30 km s$^{-1}$ (Suematsu et al. 2008; Tavabi et al. 2011, 2014a).

The widths of the TR lines place an upper limits of about ±30 km s$^{-1}$ on velocities in the TR occurring on spatial scales below the usual resolution limit of about 0"33; the resolution of velocities with IRIS is often less than 1 km s$^{-1}$. This is in agreement with the relatively small turbulence motions of TR fine structures in network rosette, typically between 1 and 3 km s$^{-1}$ (De Pontieu et al. 2014b). Sit-and-stare spectra and time-Doppler maps also indicate a TR axisymmetric spinning behavior in the elongated jet-like structures similar to spicules, and these datasets, especially in the quiet Sun regions, indicate that the integrated along the line of sight Mg II k line tend to have a bulk broad symmetric shape with red- and blue wings that masks the very small scale properties. This fact suggests twist motions of about ±30 km s$^{-1}$ within the range of spicule diameter (Pasachoff et al. 2009; Zaqarashvili and Erdlyi 2009; Tavabi et al. 2011). Time-Doppler maps within polar coronal hole indicate an extension of the TR dynamical parts of spicules known to go higher and forcing the chromosphere to be oblate (Filippov and Koutchmy 2000; Filippov et al. 2007) when over equator regions, where magnetic-field line are rather closed. Coronal hole supply mass to the fast solar wind (Tsiropoula et al. 2012; De Pontieu et al. 2014a). Magnetic forces in action seem to be the most likely source of the jet-like bright phenomena (Golub et al. 1974, 1975, 1977; Lorrain and Koutchmy 1993). The prevalence of kink and of helical Alfven and magneto-acoustic waves accompanying the guided along field lines spicules can heat the TR and the corona (Brueckner and Bartoe 1983; Tavabi et al. 2014a; Tavabi 2018).

It is known that the magnetic-field plays the most significant role at low-$\beta$,0 plasma regime, such as in TR and coronal plasma, where the charged materials are driven by magnetic forces and are confined toward the axis of flux tubes because the gas pressure decreases drastically towards the outer atmosphere layers. For an example of micro mechanism implying currents starting from the feet, see Lorrain and Koutchmy (1993, 1996, 1998). Observations have revealed that the magneto-acoustic floating modes with significant amplitudes are extremely remarkable and need to be discussed using spectral data (Martinez-Sykora et al. 2015). Variation of the parameters with the heights need to be measured.

Wave motions have been sought in the chromosphere and TR over a wide range of frequencies (Nikolskii 1965; Tavabi et al. 2015b; Khutsishvili et al. 2017). Fluctuations in Doppler velocity and intensity are also observed in the 10$^5$ K plasma. It seems to be connected to the temperature minimum layer (Georgakilas et al. 1999; Christopoulou et al. 2001) and characteristic of the cut-off frequency expected near the temperature minimum height. The magneto-sonic waves are reflected by this layer (Hollweg et al. 1982; Okamoto and De Pontieu 2011) and the mix of assumed reflected downward waves with similar periods and amplitudes of upward waves could produce the standing waves that are dominant at higher altitudes. Georgakilas et al. (1999) compared simultaneous sequences of *Ha* and He II 304 A images near the solar limb. They proposed distinguishing polar surges and giant spicules (macrospicules) among the He II structures observed beyond the solar limb. Polar surges have a complex structure when observed in Ha and an eruptive nature reminiscent of normal surges at a small scale. They found that the physical properties of the material (temperature, density, velocity) play a crucial role and, as a result, some parts of the macrospicules emit strongly in He II, whereas other parts may be better visible in *Ha* of the neutral hydrogen. It appears that limb surges become visible in He II before they do in *Ha* and remain visible for a longer period during the decay phase.

Some of these jet-like explosive events show velocities of up to 70 kms$^{-1}$ deduced from the blueshift of line profile.

These giant jets or macrospicules seem to be magnetically confined plasma as a result of interacting loops and explosive ejection in the x-null point (Yokoyama and Shibata 1995; Filippov et al. 2009, 2013). They appear to occur consecutively enough to contribute to the coronal mass supply and to the solar wind mass loss of the coronal hole regions.

Spicule-like giant features, macrospicules and surges own dramatic signatures in TR spectra made with IRIS (e.g. Tavabi et al. 2015a) and further in Figures 5 & 6. Larger velocities are often apparently seen in the so-called type II spicule as ubiquitous narrower and short life-time features that were reported from Ca H II Hinode/SOT mission broad filtergrams (De Pontieu et al. 2007). A distinguishing character of spicules is that they produce large red- and blueshifts in emission lines formed at the chromospheric and at TR levels. Their size and comparatively short life-time requires high spatial, temporal and spectral resolution, so *IRIS* rasters and slit-jaw images (SJIs) have the adequate resolution to unambiguously identify spicules and macrospicules components and are suitable to follow the time evolution of the line-of-sight velocity of limb spicules. If plasma is swaying perpendicular to the spicule axis, one may expect to see the shifts in red and blue wings frequently at the same location, and one needs to consider the plausible mechanisms causing the periodic Doppler shifts. For example, Alfven or other magneto-acoustic floating modes keep propagating along the spicules for several times. They need to be more definitely identified as Alfven or as kink waves to progress on the heating question of the TR and of the corona. Furthermore, it is suggested that the magnetic reconnection ''produce'' kinetic energy that are clues to understand the origin of explosive ejection at the magnetic null points (Yokoyama and Shibata 1995). IRIS data of very high resolution should be suitable to study their signature.

Within this work we use a unique data set with *IRIS* high spatial resolution spectra information to ascertain the nature of wave propagations. The present work provides the wave propagation at various chromospheric heights sampled by the IRIS observations at NUV wavelengths. The comparison between the different heights are performed using the time sequence spectra are showing the novel aspect of short periodic signatures of Alfven wave.

## 2. OBSERVATION AND METHOD OF REDUCTION

IRIS sit-and-stare large raster scans permitting to build up a map of the time evolution of the line-of-sight velocity of limb fine features are used. Accordingly, IRIS rasters and SJIs in the far-ultraviolet (FUV, 1389-1407 A) and near-ultraviolet (NUV, 2782-2835 Al) with the Mg II h and k resonance lines, quite similarly to the well-known H and K lines Ca II bands, are used (see Figure 1).

The raster scans are made by scanning the slit across the solar limb with a 0"33 step**.** Velocity resolution in IRIS spectra is 1 km s$^{-1}$. The IRIS NUV spectra correspond to a 0"4 resolution along the slit (Table 1) and the exposure time is 8sec, whereas the raster step size and the Nyquist criterion, assuming an excellent signal/noise (S/N) ratio, determine the spatial resolution across the slit which means twice that of step size (Pereira et al. 2014; De Pontieu et al. 2014b). The SSW/IDL program was used to stabilize the jitter effect, the value of the jitter being very small in polar regions (< 1 pixel) and higher in equatorial regions. We used level 2 data with all types of corrections. Integrated intensity maps of the Mg II k and Si IV from the very large dense raster observations are constructed from each sequence (Figure 2). Note that the resonance Mg II lines are due to a low FIP element typical of the low temperature plasma produced near the minimum temperature at the 500 km heights above $T_{500} = 1$. Their FIP is 7.65 eV.

The difficulty related to the search of Doppler velocities is known due to the, Mg II k line profile having a doublepeaks, therefore lines are actually optically thick. Being a resonance line, the classical explanation in the frame of a 1D model is to say that the outer layers of the structure are cool, therefore, an optically thick line, the radiation temperature gives information on the source function around optical depth unity. In LTE, this translates directly to the gas temperature while the source function which drops below the Planck function due to non-LTE effects. Indeed we deal with the whole extend along the line of sight of a chromospheric layer mixed with i) erupting spicular material, ii) inter- spicular material falling down, iii) a static layer of unknown origin and iv) TR material of coronal origin, including beams of electrons from reconnections, that penetrates inside this layer to a depth not yet defined and producing some evaporation in the under layers. The integration is long as it is limited just by the curvature of the layer making a shell around the Sun. We naively assume that the component(s) we study gives a signal outside the average-in-time profile, in the far wings and finally at distances outside the limb *(H > 3 arcsec )* and outside active regions (CH region) where the optical thickness is reduced (see Figure 3), note that the Alissandrakis et al. (2018) analysis shows the bulk average profile the Mg II h line is double peaked until 6" to 8" above the limb. The main reason for selecting this line is its apparent ''sensitivity" to large Doppler effects which results from its excellent efficiency to reveal for the low plasma densities of highly dynamical structures met along the line of sight, including structures situated before the main part of the chromosphere. In short, the line gives an excellent S/N ratio even for low density small features. Regarding the bulk profile, we tested to see the corresponding to the net shifts for Si IV line and the result is that the maximum intensity of Mg II h & k lines remain around 0 kms$^{-1}$, It convinced us to use this Mg II resonance spectrum line to compute the decidedly large amplitude LOS velocities to analyze the highly dynamical structures by systematically subtracting the average spectrum profile in time spectral profiles, as due to the "non- dynamical" part of the chromospheric shell (Figure 3). This is similar to the general popular method used for looking at the departure details for a bulk phenomenon that shows significant dispersion of the group average.

The C II line is from a rather high FIP element with a potential excitation corresponding to a higher temperature. The Si IV lines are produced at much higher temperatures with a potential of ionization of $Si+^2$ of 33.5 eV. In addition, slit- jaw images (SJI) with 175 x 175 $arcsec^2$ field of view (FOV) reflected off the slit/prism assembly through a filter wheel with four different filters are made available during observations. To discuss the most sensitive parameters relevant to the motion and waves analyzed off limb, we use the best S/N ratio spectra and we follow the usual procedure of extracting peculiar largely Doppler shifted features at successive heights (Vial et al. 2016). The Mg II k line profiles are now considered in details, avoiding the optically thick double peaked part by subtracting the average in time profiles for each height from the whole set of original profiles. Above the limb the maximum intensity of the Mg II k line remains around 0 kms$^{-1}$ (Alissandrakis et al. 2017). The method gives good results well above the limb, which means in the far wings with presumably small optical thicknesses even for heights under 8".

Note that under 8", the continuity apparent in the behavior of the deduced profiles, see Figures 5 and 6, gives hope that we are not far from the real behavior of the 'individual' analyzed structure met along the line of sight when a well coherent signature appears. The resulting subtracted spectra are further call simply "d-spectra" to make it simpler (d means difference).

## 3. RESULTS AND INTERPRETATIONS

Doppler maps (Figure 4), intensity map (Figure 2) and time-spectra's diagrams (Figures 5 and 6) in Mg II k line of IRIS observations consistently illustrate the occurrence of 3-4 minute oscillations. Time integrated spectral profiles show an explosive event where a macrospicule appears at the beginning of the time series. The maximum value of Doppler velocity on this macrospicule is reached in the blue wing (note left-hand side in Figures 2 and 3 of right panel).

The wave motions are assumed to be present in only two directions of propagation: in azimuthal and in longitudinal directions of the spicule axis. The azimuthal motions are interpreted as Alfvenic waves. These waves could be included in all the transverse displacements such as kink (m = 1), swaying motions *etc*. In the first approximation, the lowest modes of these waves consist of similar patterns.

The TR time-Doppler and time-spectra's diagrams for very large sit-and-stare rasters are remarkable in showing the presumably axisymmetric rapid red- and successive blueshifts, which indicate a general torsional motions that propagate along the spicule axis at an average speed of ±40 kms$^{-1}$ (De Pontieu et al. 2014a; Tavabi et al. 2015b). The amplitudes of the rotational flows are comparable to the spicule swaying motions, so it is likely that the Alfvenic waves propagate along the spicules, as reported in the literature (Tavabi et al. 2014a,b; De Pontieu et al. 2014a; Kitiashvili et al. 2013; Gonzalez-Aviles et al. 2017). Sometimes, this has been interpreted as spinning motions in doublet threads or multi-components spicules (Suematsu et al. 2008; Tavabi et al. 2011). The wavelet analysis see Figure 7, often shows a periodic oscillations of the low-frequency of magneto-sonic waves at 3-4 mins and evidence of a much high-frequency

**Table 1.** Dataset properties

| date, OBSID | Time [UT] | Target X, Y [arcsec] | Slit length | Spectral cadence, exp. time [sec.] | SJI, (FOV) | SJI, Cadence [sec.] | SJI bands [A] |
|---|---|---|---|---|---|---|---|
| 2013-10-16, 3820009402 | 21:40-22:28 | [688 698] | 60" | 9, 7.9 | 60" x 65" | 18 | 1400 & 2796 |
| 2014-01-27, 3800009253 | 11:59-14:48 | [737 629] | 119" | 10, 7.9 | 119" x 119" | 19 | 1330 & 1400 |
| 2014-02-19, 3800258253 | 16:27-17:23 | [6 -980] | 119" | 9, 7.9 | 119" x 119" | 23 and 19 | 1330 & 1400 & 2832 |

oscillatory behavior in red and blueshifts with periods as low as 30 second.

The behavior along the slit needs a special discussion (see Figures 8, 9 and the online movie). The concept of multi-component flows is another interesting aspect seen in the IRIS time-Doppler maps which makes the discussion rather difficult until more work is done on the subject. The heating mechanisms presumably operate in the TR and the corona (De Pontieu et al. 2004; Rouppe van der Voort et al. 2015).

The 3-4 minute magneto-acoustic waves are also presented in the time-altitude variations of the propagating waves demonstrated using time sequence of spectra of several adjacent layers (see Figures 5, 6 and 8 and online available movie). Wavelet power spectra of Figure 7 indicate a drastic decrease in the 3-4 minute oscillation powers at higher altitudes and an increase in the power of high-frequency oscillations (down to 20 to 30 second periods) in upper layers. For the first time, the well identified 100 sec period presumably Alfven waves at the height of TR are seen in the Mg II k spectral line: see in Figures 5, 6, 7 and 9. The limb positions for the sequence of February 19, 2014 are particular as it is in the South Pole, so the solar rotation motion produces a negligible effect. In this dataset, the slit is located perpendicular to the limb at the South Pole (Figure 1, bottom snapshots).

By comparing the spectrally resolved time sequence of spectra of Mg II k at various heights above the limb, a rough estimate of the periods, amplitudes, and position in the higher layers of TR are estimated. For this, we use a time sequence of d-spectra from the near- to off-limb altitudes, where we still have enough signal-to-noise ratio to make plots for the dynamical components (Figures 5, 6 & 8, and see the online available movie). Apart from this, the panels (the limb corresponds to the top of the panel) reveal that all initial oscillatory features appear with a decreased intensity at higher layers. This is a possible evidence of ions dragged by an initial upward waves reach all the way up to the TR with the same periods and similar amplitudes in our case.

The overlapping effects are more important for smaller features that are in larger number and for shorter periods that we rather avoid to discuss, as with Figure 8.

## 4. DISCUSSION

We note: a dominant peak of power spectrum around 3-4 minute. This peak seems to correspond to magneto-sonic waves mainly appearing near the limb, and disappearing at higher levels. Spicules in the polar coronal hole region are more or less vertical and they are more inclined in the equator regions where the magnetic-field lines are closed. The inclination could help the mechanical acoustic waves to penetrate and propagate the volume of the corona more easily (De Pontieu et al. 2004).

High frequency atypical oscillations are seen in Figures 5, 6, 8 and 9 with periods around 100 sec. that were discovered using the SOT *(Hinode)* Ca II H filtergram series (Tavabi et al. 2011). Evidence of pulse- like displacements appears suggesting rather violent processing well above the limb as illustrated in Figure 8.

They need a special attention because we believe they could explain the heating of the corona higher up into the corona. We identify these waves with Alfven waves when no correlation exists between the large Doppler shifts we measured and the brightness variations that would reveal a possible variation of the pressure as it would be the case for a kink wave. In addition, such high frequency wave propagates better in the hot corona where the sound speed is also higher. We tried to measure some phase velocity using the analysis of Figure 7, without success but an evaluation of the lower value points to $> 500$ kms$^{-1}$. Accordingly, the period of the best observed fast waves is of order of 100 second and they dominate at higher altitudes. This period could be related to the spinning motions of spicules.

The time-Doppler maps show a dominating spinning motion patterns with an average helical velocity in the range of $\pm 30$ kms$^{-1}$. With the assumption that there is not any discontinuity, except for the spatial and temporal scales both of which are less than the spicule diameter, scale height, and life-time, it certainly indicates that these twisty motions are attached to the individual structures. It means that a bilateral velocity components originate at the same location without any time, space and spectral discontinuities in time-Doppler plots. One may interpret these components as the components of a rotating spicule.

Twisty motions have already been reported with a velocity of about 25 kms$^{-1}$ (De Pontieu et al. 2014a; Tavabi et al. 2015b) and they are predicted by magnetic vortex plasma simulation (Kitiashvili et al. 2013; Gonzalez-Aviles et al. 2017). An explosive event could eject plasma in all directions and produce equal red- and blueshifts, as for a bi-directional jets (Tavabi et al. 2015a). The asymmetric (with a tilt angle) profiles on the red and blue wing of the events could be explained by a rotational ejection. Spinning configurations are clearly suggested in the time-Doppler maps we produced. They were also described in the literature as Alfvenic waves propagation (Brueckner and Bartoe 1983; Martinez-Sykora et al. 2015).

The raster spectral lines and SJI's observations suggest that giant spicules (or miniature surges) are also impulsive phenomena and much more numerous on a scale smaller than polar surges. They give the impression of jet-like features, similar to ordinary spicules except for their dimensions. Thus, they appear to be as result of magnetic reconnection on a scale smaller than that of polar surges, revealing the fact that giant spicules protrude higher into the hot corona (Georgakilas et al. 1999). Hence, these observations suggest that there is a close association between polar surges with explosive events, supporting the hypothesis that magnetic reconnection triggered by emerging flux provides the acceleration mechanism for this polar region event. Furthermore, the equator region sequence (October 16, 2013) shows a bilateral motion in symmetric schema in near- and off-limb positions. The Doppler velocities in Figures 7 & 8 are approximately equal in both wings with typical peaks in $\pm 30$ kms$^{-1}$; in these figures the left panels show features that are often normal to the dispersion direction. Finally let us notice that the importance of high frequency waves should not be underestimated in the frame of the heating problem of the corona where these waves, including pure acoustic waves, propagate much more easily than the popular 5 and/or 3 min oscillations. Even more significantly for evaluating observations, their amplitudes are indeed greatly underestimated because no account is made of the instrumental attenuation at high frequencies occurring for small scale variations (under the spatial resolution, they are just removed) and even more, the superposition effects along the line of sight are becoming more important with the cross- section of structures getting smaller and accordingly, the number of integrated features is getting larger but the resulting contrast drastically decreases. Just to mention one important typical geometric effect let us point out that along the line of sight we have a lot of overlapping of different optically thin and thick dynamical features impossible to properly take into account. Just to mention one important typically geometric effect remember that along the line of sight we have a lot of overlapping of different optically thin and optically thick dynamical features impossible to properly take into account. With a large coronagraph looking at the profiles of the Fe XIV coronal line, high frequency (with dominant periods near 85 and 43 sec) transverse velocity oscillations were already observed in Doppler shifts in the low-,0 corona, (see e.g. Koutchmy et al. 1983) above regions showing episodic small surges or micro-surge in $H\alpha$, suggesting that magnetic waves propagating towards the corona are produced as described here. Thanks to the high resolution provided by IRIS we were able to show that plenty of high frequency waves are produced in the low corona suggesting that the so- called theoretically suggested turbulent heating could be a reality. Eventually, the presented material confirms that Alfven waves are produced at the feet of the field lines and they propagate far away into the corona. More work should be done on fast and small scale variations inside the thick multi- component shell surrounding the Sun, including the analysis of hotter coronal lines.

**Acknowledgments** We warmly acknowledge the work of our referee, who provided an extended detailed report and added many interesting suggestions and requestsfor greatly improving the paper. The authors gratefully acknowledge the use of data from the IRIS databases. IRIS is a NASA small explorer mission developed and operated by LMSAL with mission operations executed at NASA ARC with the contribution of NSC (Norway). We thank B. Filippov (Pushkov Moscow Institute) and J-C. Vial (IAS Orsay) for meaningful discussions.

## REFERENCES

Alissandrakis, C. E.; Zachariadis, Th.; Gontikakis, C., 2005, Proceedings of the 11th European Solar Physics Meeting "The Dynamic Sun: Challenges for Theory and Observations" (ESA SP-600). 11-16 September 2005, Leuven, Belgium. Editors: D. Danesy, S. Poedts, A. De Groof and J. Andries. Published on CDROM., id.54.1.

Alissandrakis, C. E., Koukras, A., Patsourakos, S., & Nindos, A. 2017, A&A, 603, A95.

Alissandrakis, C. E., Vial, J.-C., Koukras, A., Buchlin, E., & Chane-Yook, M. 2018, SoPh, 293, 20.

Brueckner, G. E. and Bartoe, J.-D. F. 1983, ApJ, 272, 329.

Christopoulou, E.B., Georgakilas, A.A., and Koutchmy, S., 2001, SoPh., 199, 61.

De Pontieu, B., Erdelyi, R., & James, S. P. 2004, Nature, 430, 536.

De Pontieu, B. et al. 2007, PASJ, 59, S655.

De Pontieu, B., Rouppe van der Voort, L., McIntosh, S. W., et al. 2014a, Science, 346, 1255732
De Pontieu, B., Title, A. M., and Lemen, J. R., et al. 2014b, SoPh, 289, 2733.
Filippov, B., & Koutchmy, S. 2000, SoPh, 196, 311
Filippov, B., Koutchmy, S. & Vilinga, J. 2007, A&A, 464,1119
Filippov, B., Golub, L., Koutchmy, S.: 2009, SoPh, 254, 259. DOI: 10.1007/s11207-008-9305-6.
Filippov, B., Koutchmy, S., Tavabi, E.: 2013, SoPh, 286, 143. DOI: 10.1007/s11207-011-9911-6.
Georgakilas, A. A., Koutchmy, S., Alissandrakis, C.F.: 1999, A&A 341, 610.
Golub, L. Krieger, A.S. Silk, J.K. Timothy, A.F. and Vaiana, G.S. 1974, ApJ, 189, L93.
Golub, L., Krieger, A. S., and Vaiana, G. S. 1975, SoPh, 42, 131.
Golub, L., Krieger, A. S., Harvey, J. W., & Vaiana, G. S. 1977, SoPh, 53, 111.
Gonzalez-Aviles, J. J., Guzman, F. S., & Fedun, V. 2017, ApJ, 836, 24.
Hollweg J. V., Jackson S., Galloway D., 1982, SoPh, 75, 35
Kitiashvili, I. N., Kosovichev, A. G., Lele, S. K., Mansour, N. N., & Wray, A. A. 2013, ApJ, 770, 37.
Khutsishvili, D.; Zaqarashvili, T. V.; Khutsishvili, E.; Kvernadze, T.; Kulidzanishvili, V.; Kakhiani, V.; Sikharulidze, M., 2017, A&SS, 362, 10.
Martinez-Sykora, J., Rouppe van der Voort, L., Carlsson, M., et al. 2015, ApJ, 803, 44.
Koutchmy, S., Zhugzhda, I. D., & Locans, V. 1983, A&A, 120, 185
Lorrain, P. and Koutchmy, S.1993, A&A, 269, 518
Lorrain, P. and Koutchmy, S. 1996, SoPh, 165, 115
Lorrain, P. and Koutchmy, S. 1998, SoPh, 178, 39
Nikolskii, G. M. 1965, Astron. Zh. 42, 86
Okamoto, T. J. and De Pontieu B. 2011, Astrophys. J. Lett. 736, L24
Pereira, T. M. D., De Pontieu, B., & Carlsson, M. et al. 2014, ApJ, 792, L15
Pasachoff, J. M., Jacobson, W. A., and Sterling, A. C., 2009, SoPh, 260, 59.
Rouppe van der Voort, L., De Pontieu, B., Pereira, T. M. D., Carlsson, M., & Hansteen, V. 2015, ApJ, 799, L3
Suematsu, Y., Ichimoto, K., Katsukawa, Y., et al. 2008, in Astronomical Society of the Pacific Conference Series, Vol. 397, First Results From Hinode, ed. S. A. Matthews, J. M. Davis, & L. K. Harra, 27
Tavabi, E., Koutchmy, S., Ajabshirzadeh, A., 2011, New Astron. 16, 296. DOI: 10.1016/j.newast.2010.11.005.
Tavabi, E.: 2014a, Ap&SS, 350, 489. DOI: 10.1007/s10509-014-1807-0.
Tavabi E., 2014b, Ap&SS, 352, 43. DOI 10.1007/s10509-014-1807-0.
Tavabi, E., Koutchmy, S., and Golub, L. 2015a, SoPh, 290, 2871.
Tavabi, E., Koutchmy, S., et al.: 2015b, Astron. Astrophys, 573, 7. DOI: 10.1051/0004-6361/201423385.
Tavabi, E.: 2018, MNRAS, 486, 868. DOI: 10.1093/mnras/sty020.
Vial, J.-C., Pelouze, G., Heinzel, P., Kleint, L., Anzer, U.: 2016, SoPh. 291, 67, DOI: 10.1007/s11207-015-0820-y.
Tsiropoula, G., Tziotziou, K., and Kontogiannis, I., et al. 2012, Space Sci. Rev. 169, 181.
Yokoyama, T., & Shibata, K. 1995, Nature, 375, 42.
Zaqarashvili, T. V., & Erdelyi, R. 2009, Space Sci. Rev., 149, 355
Zirin, H. 1996, SoPh., 169, 313.

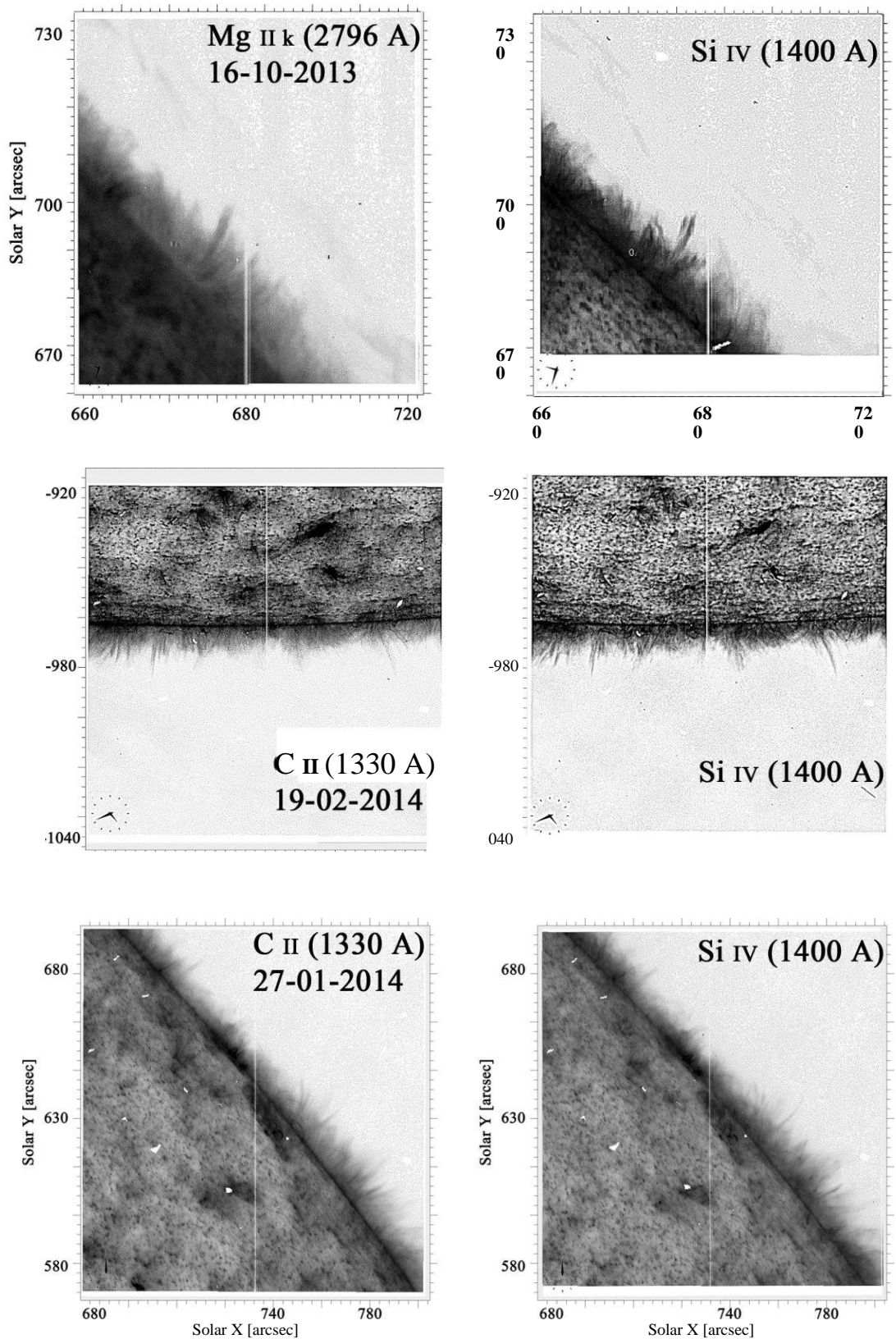

**Figure 1.** Negative selected frames extracted from the sequences of October 16, 2013 (top), February 19, 2014 (middle pannel) and taken 27 January 2014 (bottom) shown in negative to improve the visibility. Some unsharp masking filtering is used for the CH image in the middle panel. The slit is seen as a vertical long thin line perpendicular to the limb on the second dataset taken on February 19, 2014.

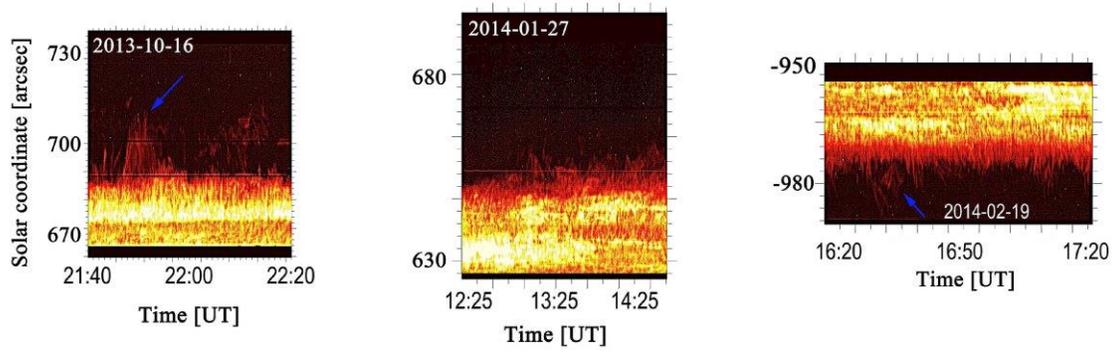

**Figure 2.** Time-intensity maps of the sit-and-stare large rasters according to the summarized properties of observations in Table 1 and also shown in Figure 4, the integrated wavelength range is in order of 1 A between 2795.9 to 2796.9 A

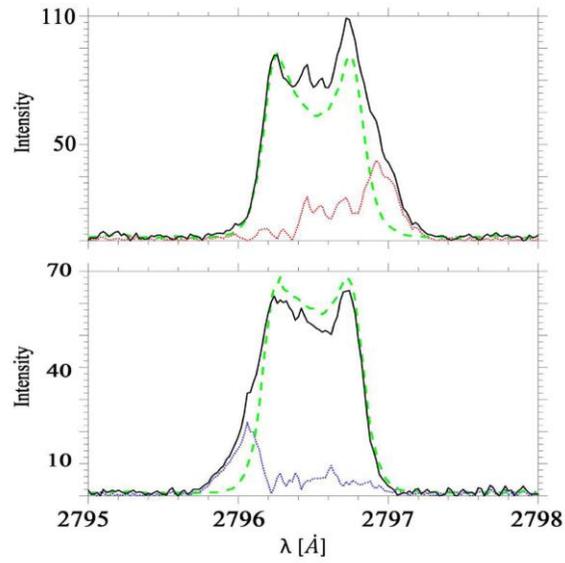

**Figure 3.** Examples of a red shift of Mg II k lines (dotted red curve) resulting from the subtraction of time-averaged profile (in dashed green curve) from the original profile (solid black curve) for a selected frame of February 19, 2014 at 17:30:53 UT (top). The bottom panel is showing the a blue shift for this dataset at 16:32:39 UT, at the position about 3"3 above the limb for the red shift and 3"7 for the blue shift.

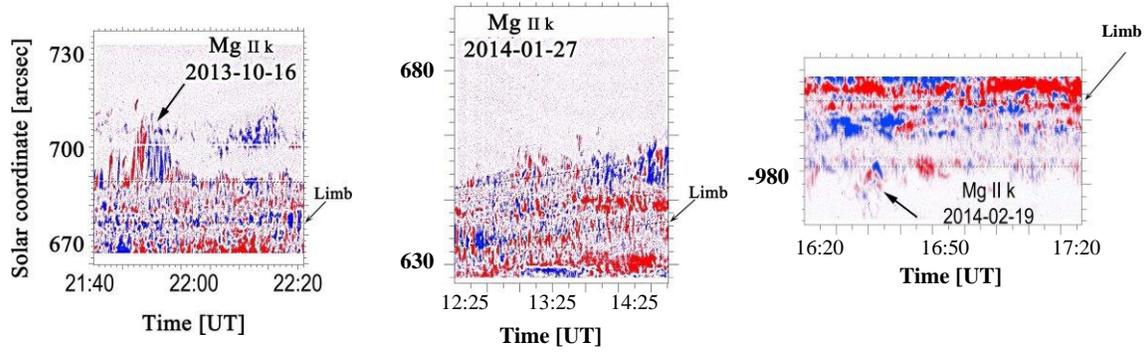

**Figure 4.** Time- Doppler maps formed by subtracting blue- and red-shifted intensities in the wings of Mg II k line (2796.32 $A$ core) of sit-and-stare large and very large dense rasters. The color scale is saturated at [-30 +30] km s$^{-1}$ corresponding to ±0.3A. Horizontal dotted lines show the boundaries of layers with respect to the limb shown. The elongated structures which are significantly shifted in both wings (right panel) are parallel and adjacent to each other, and are repeated over the whole duration of the observations. Limb are shown by arrow and the inserted arrows showing the macrospicules for the first and the last time sequence. The positions of the limb are shown and the arrows point to the macro- spicule events observed in 2 instances.

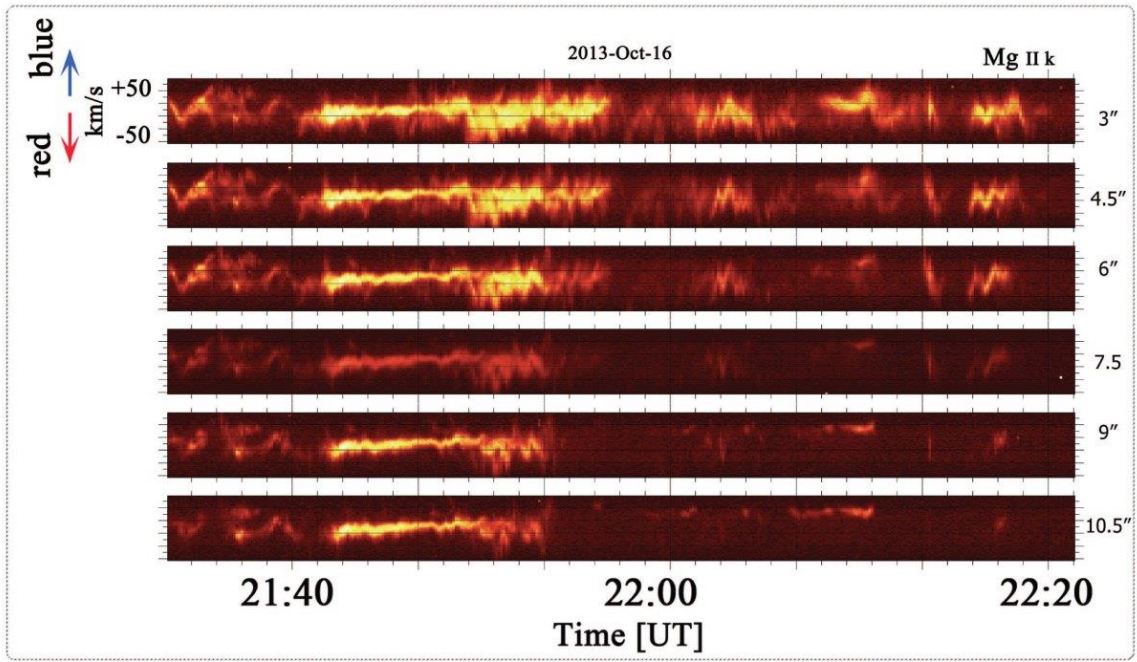

**Figure 5.** Time sequence of d-spectra in Mg II k line of Oct. 16, 2013 in adjacent layers above the limb on the quiet Sun region at equator. The height difference between each plot is ~ 1.5 arcsec. Waves with periods near 100 sec are well detected near 21:52 to 21:56 UT.

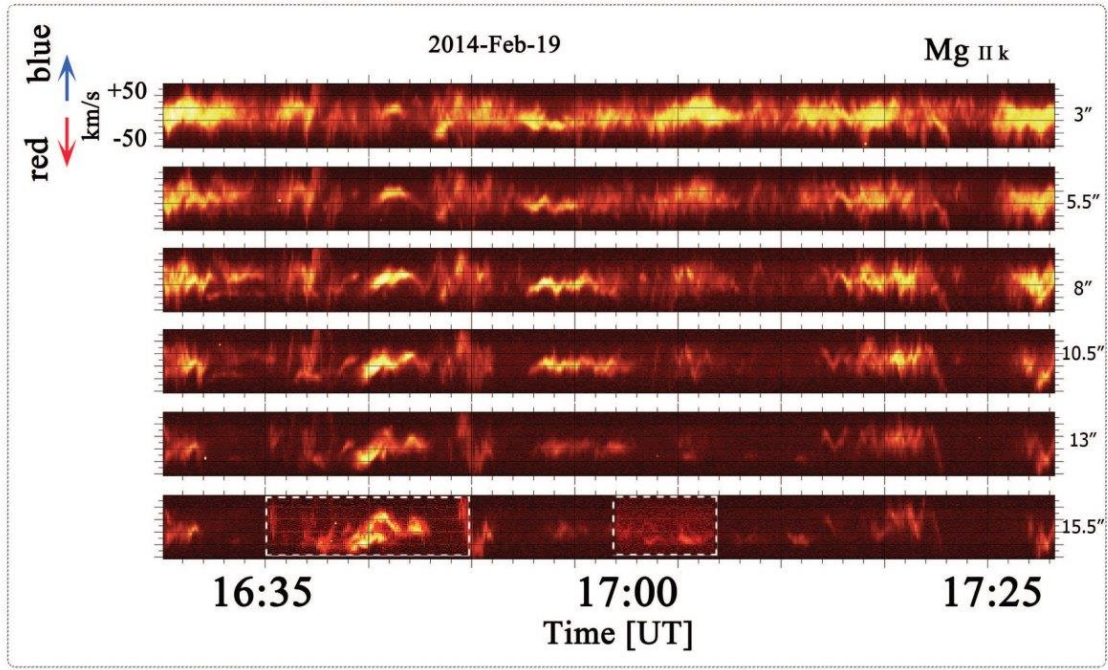

Figure 6. Time sequence of spectra d- spectra in Mg II k line of Feb. 19, 2014 in adjacent layers above the limb at the polar coronal hole. The height difference between each plot is ~ 2.5 arcsec. 100 sec period waves are seen at the beginning of the sequence and also around 17:23 UT.

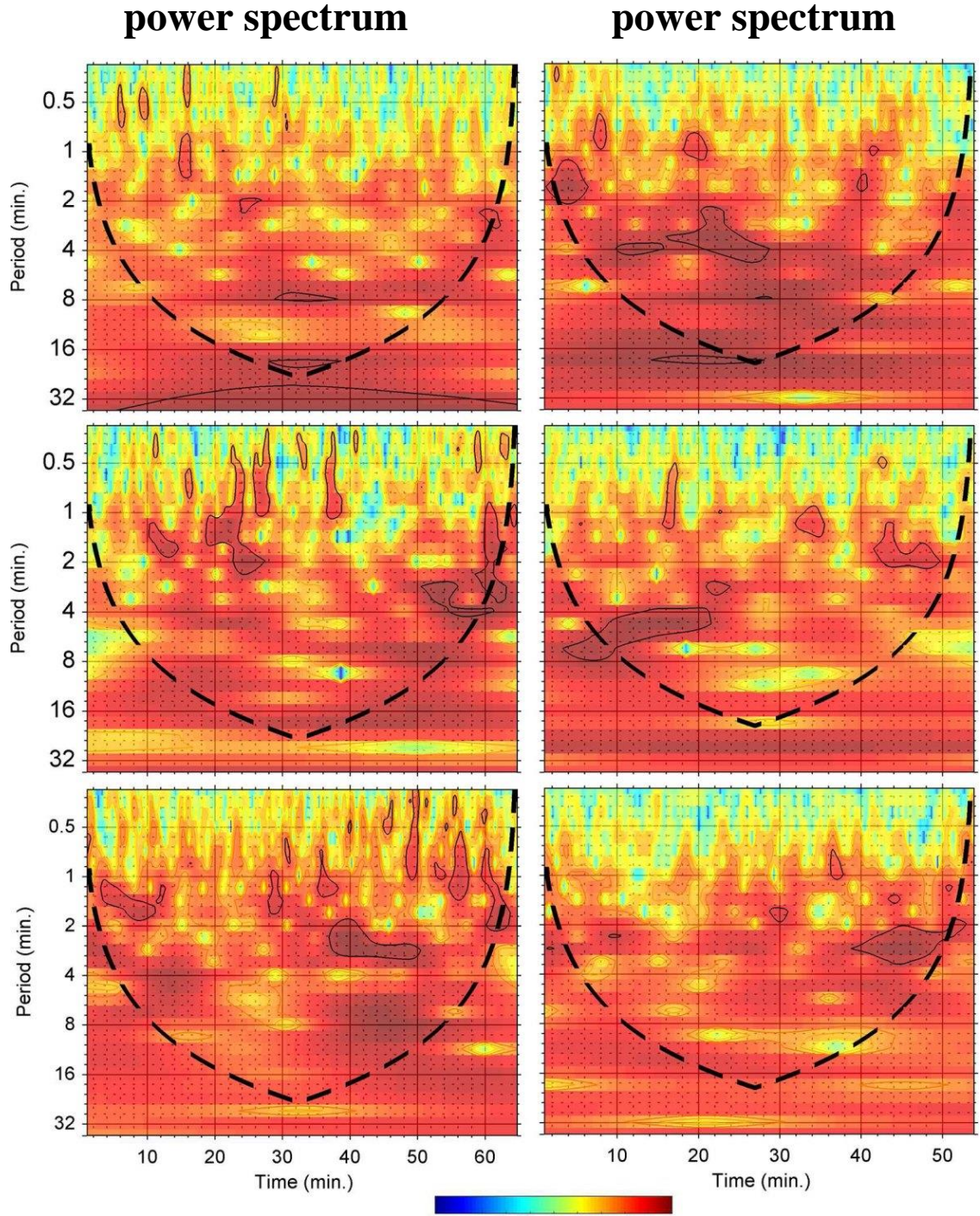

**Figure 7.** Wavelet power spectra powers of the oscillations in Mg II k in d- spectra for three layers above the limb (from bottom to top they are correspond to the lowest level towards the highest level), with 4 arcsec difference distance for October 16, 2013 observations in the left panels correspond of the near equatorial region. In the coronal hole region dataset (February 19, 2014), the distance separation is 6 arcsec (right panels). The contours determine the 95% confidence level which was calculated by assuming a white noise background spectrum. Wavelet transform suffers from the wraparound errors at both edges of a time series (of finite length). The regions in which these effects are important are defined by the cone of influence. A horizontal colorbar below showing the power of wavelet spectrum, the redmost area corresponding to the strongest spectrum peaks.

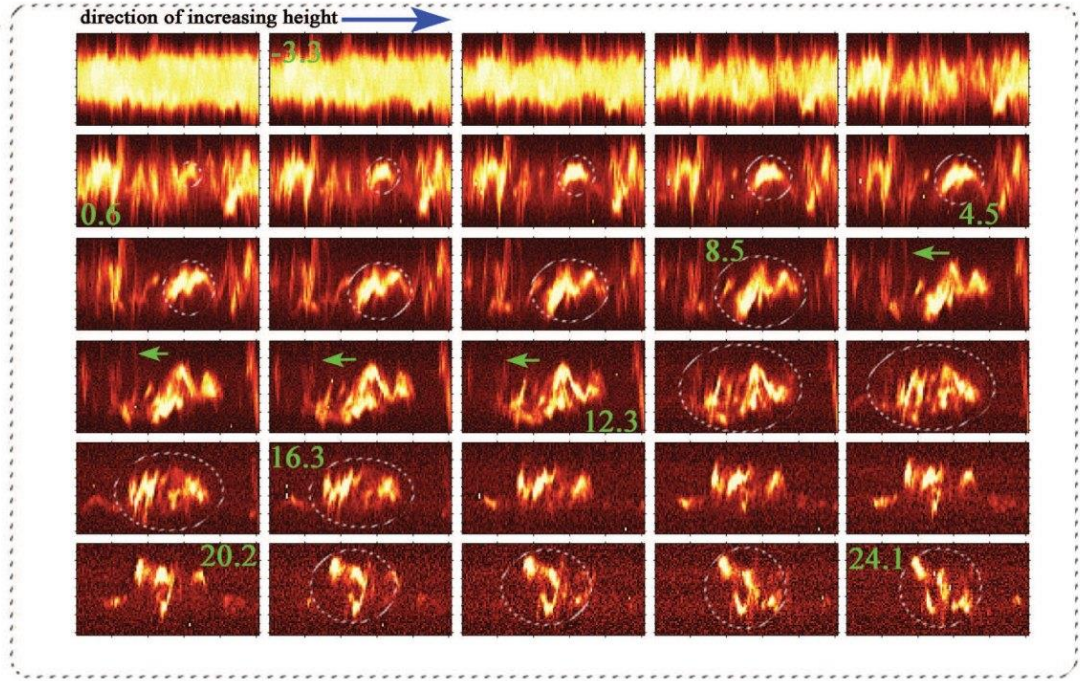

**Figure 8.** Time sequence of Mg II k spectra for the region that was in Figure 7 (the first dashed box at bottom) for Feb. 19, 2014. The time laps of each panel is ~ 15 minutes between 16:35 to 16:50 UT, and the height distance between successive panels is 3 pixels or 1 arcsec (pixel's size is ~0"33) in space. The green arrows a peculiar feature. The approximate heights are inserted in some frames with an error of about ±0"64 arcsec (see the online available movie for higher resolution in space). Note between 14" and 18" the evidence of waves with very short periods (<1 min). Between 5" and 10" the period of the main wave is around 3 min.

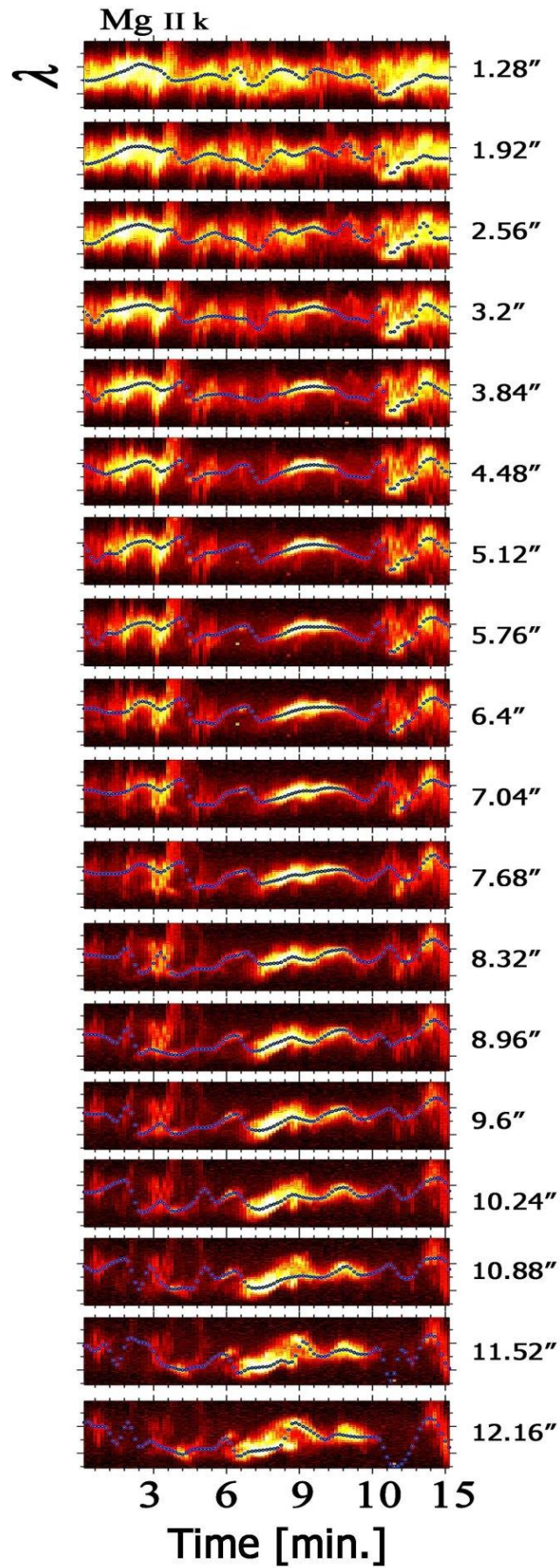

**Figure 9.** Extracted time sequence of spectra [1"3 ,12"1] in Mg II k band with a possible reversal of Doppler velocities with heights for Feb. 19, 2014 and time interval is the same as Figure 8. At left note the wave with a 100 sec period at 3 min time that does not show a measurable phase shift with the heights in this interval.